\documentclass[twocolumn]{webofc}
\usepackage[varg]{txfonts}   
\begin{document}
\title{A walk along the proton drip-line by $\beta$-decay spectroscopy}
%

\author{\firstname{Sonja E. A.} \lastname{Orrigo}\inst{1}\fnsep\thanks{\email{Sonja.Orrigo@ific.uv.es}} \and
        \firstname{Berta} \lastname{Rubio}\inst{1} \and
        \firstname{William} \lastname{Gelletly}\inst{2}
}

\institute{Instituto de F{\'i}sica Corpuscular, CSIC-Universidad de Valencia, E-46071 Valencia, Spain 
\and
           Department of Physics, University of Surrey, Guildford GU2 7XH, Surrey, UK 
          }

\abstract{%
During the last decade we have carried out a systematic study of the $\beta$ decay of neutron-deficient nuclei, providing rich spectroscopic information of importance for both nuclear structure and nuclear astrophysics. We present an overview of the most relevant achievements, including the discovery of a new exotic decay mode in the \textit{fp}-shell, the $\beta$-delayed $\gamma$-proton decay in $^{56}$Zn, the first observation of the 2$^+$ isomer in $^{52}$Co and the latest results on the heavier systems $^{60}$Ge and $^{62}$Ge. We also report on our deduced mass excesses in comparison with systematics and a recent measurement. Finally, we summarise our results on the half-lives of $T_z=$ -1/2, -1 and -2 neutron-deficient nuclides, and analyse their trend.
}
\maketitle
\section{Introduction}
\label{intro}
The investigation of nuclei close to the limits of nuclear stability, \textit{exotic nuclei}, is at the frontier of modern nuclear physics. Their study is challenging but it is becoming more possible thanks to new-generation facilities for the production and acceleration of radioactive ion beams (RIBs). Neutron-deficient nuclei can be populated even up to the boundaries of nuclear stability, the \textit{proton drip-line}, thus enabling one to perform detailed decay studies. 

Decay-spectroscopy experiments provide us with structure information of paramount importance \mbox{\cite{Orrigo2014, Orrigo2016, OrrigoPRC2, Kucuk2017, Orrigo2018, Orrigo2021}}, such as the half-life of the unstable nucleus and the energies and branching ratios for $\beta$-delayed $\gamma$ or particle emission, e.g. protons, $\alpha$ particles or neutrons, depending on whether the $\beta$-decaying nucleus is either neutron-deficient or neutron rich. This is important since it allows us to determine the energy levels populated in the daughter nucleus together with their $\beta$-feedings and thus reconstruct the partial decay scheme. From all this information one can finally determine the absolute values of the Fermi \textit{B}(F) and Gamow-Teller \textit{B}(GT) transition strengths and, in favourable cases, also deduce the mass excesses of the daughter and parent nuclei.

If isospin were a good quantum number in nuclear physics then mirror nuclei, where the number of neutrons and protons are exchanged, would be identical. In reality their properties, such as the structure of the levels, quantum numbers, half-lives, $\beta$-decay strengths, etc., are very similar but not identical. There can be differences which break the ideal isospin symmetry. Charge-exchange (CE) reactions are the mirror strong interaction process of $\beta$ decay \cite{Fujita2005, Fujita2011}. The comparison between $\beta$-decay data and CE reactions carried out on the stable mirror target allows us to investigate fundamental questions related to the role of isospin in atomic nuclei.

Moreover, many exotic nuclei lie on the reaction pathways involved in various processes of nucleosynthesis, responsible for the production of the chemical elements in the Universe. Nuclear-structure properties such as half-lives, masses and $\beta$-strengths are of great significance for nuclear astrophysics. One such nucleosynthesis processes is the rapid proton-capture, \textit{rp-process}, happening in explosive stellar environments \cite{Sch98}. It acts on the neutron-deficient side of the nuclear chart and produces many medium-mass/heavy proton-rich elements, passing through neutron-deficient nuclei in the \textit{fp}-shell and above.

\begin{figure*}[ht]
\centering
\includegraphics[width=2\columnwidth]{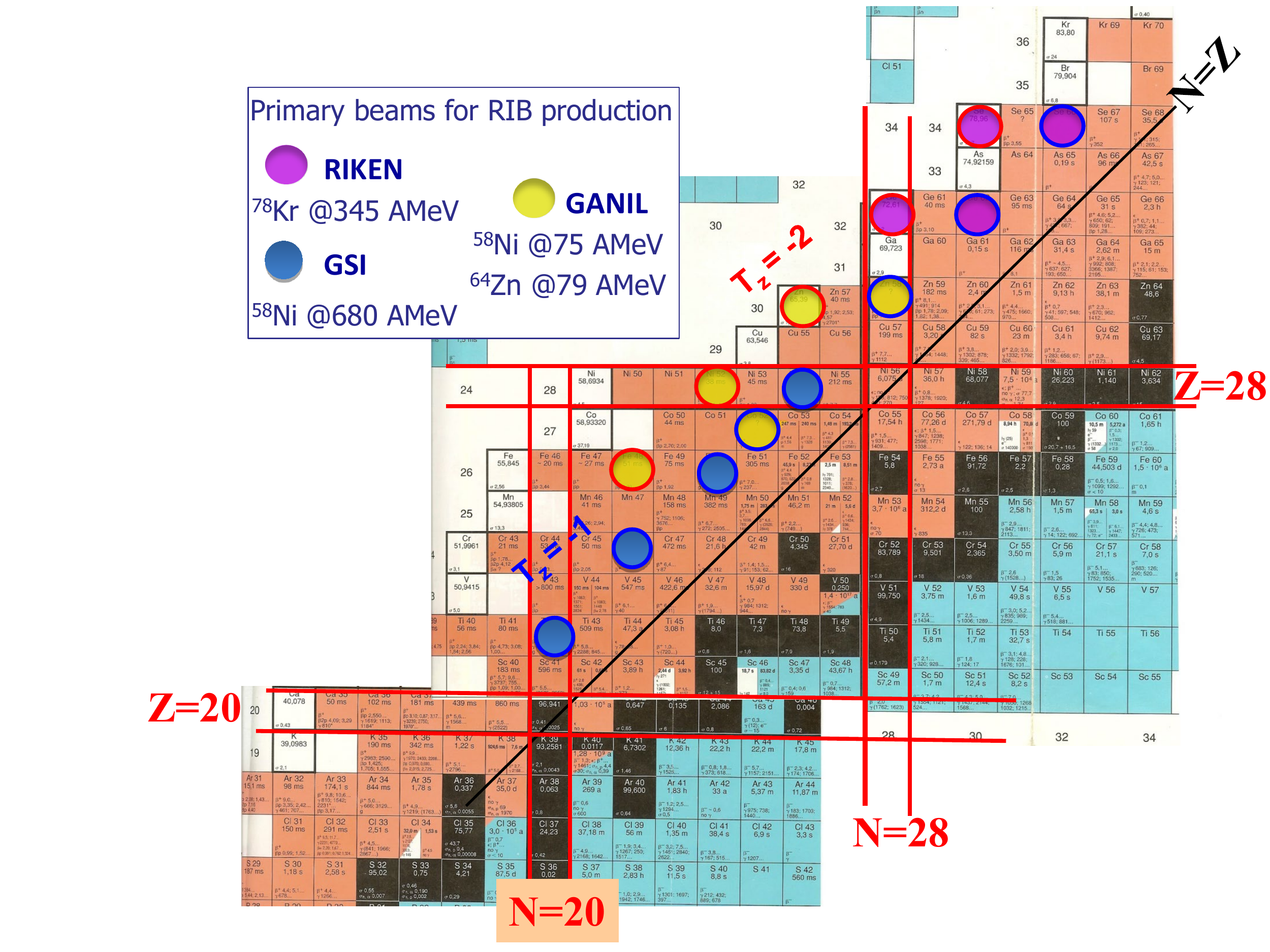}
\caption{Summary of the $\beta$-decay spectroscopy experiments carried out along the proton drip-line, in the \textit{fp}-shell and above \mbox{\cite{Orrigo2014, Orrigo2016, OrrigoPRC2, Kucuk2017, Orrigo2018, Orrigo2021, Molina2015}}. Neutron-deficient nuclei produced at the GSI/GANIL/RIKEN laboratories are marked by a blue/yellow/purple circle, where a blue or red 
circumference line indicates a value of $T_z=$ -1 or -2. The primary beams used in the different laboratories and their energies are reported in the textbox.}
\label{fig1}
\end{figure*}

During the last decade we have performed a systematic study of neutron-deficient nuclei along or close to the proton drip-line by $\beta$-decay spectroscopy experiments with implanted RIBs. Figure~\ref{fig1} shows the nuclei studied, spanning the \textit{fp}-shell and above: the nuclei marked with blue/yellow/purple circle where produced at the GSI/GANIL/RIKEN laboratories, respectively, while a blue or red 
circumference line indicates a value of -1 or -2 for the third component of the isospin quantum number, $T_z$. The textbox indicates the primary beam that was used to produce the secondary RIB of interest in the various laboratories, where in the case of GANIL a $^{58}$Ni primary beam was used for the experiment focused on the production of the $T_z=$ -2, $^{56}$Zn nucleus \cite{Orrigo2014, Orrigo2016}, while a $^{64}$Zn beam was used for the experiment devoted to the study of the $T_z=$ -1, $^{58}$Zn nucleus \cite{Kucuk2017}. The results from the GSI experiment are published in Ref. \cite{Molina2015}. Results from the GANIL experiments are published in Refs. \cite{Orrigo2014, Orrigo2016, OrrigoPRC2, Orrigo2018, Kucuk2017}. The most recent study of the $\beta$ decay of $^{60}$Ge and $^{62}$Ge, performed at RIKEN, is reported in Ref. \cite{Orrigo2021}. Results on $^{64}$Se and $^{66}$Se will be published soon.

\begin{table*}[h!]
\centering
\caption{Nuclei produced at GANIL and RIKEN: reference, isotope of interest, its $T_z$ value and the experimental results for the number of implants $N_{imp}$ and half-life $T_{1/2}$. The last two columns specify the laboratory and primary beam used for their production.}
\label{tab-1}
\begin{tabular}{lllllll}
\hline
Ref. & Isotope & $T_z$ & $N_{imp}$ & $T_{1/2}$(ms) & Laboratory & Beam, energy(AMeV) \\\hline \\[-0.35cm] 
\cite{Orrigo2014} & $^{56}$Zn & -2 & $8.9\times 10^3$ & 32.9(8) & GANIL & $^{58}$Ni, 75 \\
\cite{Orrigo2016} & $^{48}$Fe & -2 & $5.0\times 10^4$ & 51(3) &  & \\
\cite{Orrigo2016} & $^{52}$Ni & -2 & $5.3\times 10^5$ & 42.8(3) &  & \\
\cite{OrrigoPRC2} & $^{52}$Co$^m$ & -1 & $2.3\times 10^5$ $^{(a)}$ & 102(6) &  & \\
\cite{OrrigoPRC2} & $^{52}$Co$_{g.s.}$ & -1 & $4.1\times 10^6$ $^{(b)}$ & 112(3) &  & \\
\cite{Kucuk2017}  & $^{58}$Zn & -1 & $1.8\times 10^5$ & 86(2) &  & $^{64}$Zn, 79 \\\hline \\[-0.35cm]
\cite{Orrigo2021} & $^{62}$Ge & -1 & $2.1\times 10^6$ & 73.5(1) & RIKEN & $^{78}$Kr, 345 \\
\cite{Orrigo2021} & $^{60}$Ge & -2 & $1.5\times 10^4$ & 25.0(3) &  & \\\hline
\end{tabular}
\\ \raggedright{$^{(a)}$ Estimated based on production as a by-product of $^{52}$Ni implantation \cite{Orrigo2016, OrrigoPRC2}. \\ $^{(b)}$ Calculated as 83\% of the total $^{52}$Co implants ($5.0\times 10^6$) including both g.s. and isomeric state \cite{OrrigoPRC2}.}
\end{table*}

In the following we focus especially on the GANIL and RIKEN experiments. The experimental setups and the results obtained are discussed in Sect.~\ref{sec1}. Our results on the mass excesses are summarised in Sect.~\ref{sec2} and compared with the mass evaluation systematics and a recent measurement \cite{Paul2021}. Finally, half-life trends for $T_z=$ -1/2, -1 and -2 nuclei, including all the newly measured values, are presented in Sect.~\ref{sec3}. 

\section{$\beta$-decay spectroscopy experiments and results}
\label{sec1}
As shown in figure~\ref{fig1}, our $\beta$-decay spectroscopy experiments were performed worldwide at different RIB facilities, but conceptually they are the same. A primary beam is accelerated and fragmented on a thick target: for example beams of $^{58}$Ni or $^{64}$Zn are fragmented on a natural Ni target at GANIL, while $^{78}$Kr is fragmented on a Be target at RIKEN. The fragments are then selected by the LISE3 separator \cite{Anne1992276} at GANIL, or by the BigRIPS separator \cite{Fukuda2013} at RIKEN. Thereafter they are implanted into Double-Sided Silicon Strip Detectors (DSSSD), used to detect both the implanted heavy ions and any subsequent charged-particle decay: a 300-$\mu$m thick DSSSD at GANIL, or the WAS3ABi setup \cite{Nishimura2012} comprising three 1-mm thick DSSSDs at RIKEN. Finally, arrays of high-purity Ge detectors surround the DSSSD setup to detect the $\beta$-delayed $\gamma$-rays: four EXOGAM clovers \cite{exogam} at GANIL, or the EURICA array \cite{Soderstrom2013} consisting of 12 clusters at RIKEN. More details on the setups, the procedures employed for particle identification and data analysis are available in Refs. \cite{Orrigo2016, Orrigo2021}.

Table~\ref{tab-1} shows the number of implanted ions ($N_{imp}$) together with their measured half-lives $T_{1/2}$ and details, such as the laboratory and primary beam, of their production. The unprecedented statistics available at the RIKEN Nishina Center allowed us to extend the systematic exploration of neutron-deficient nuclei to higher masses, along the proton drip-line.

Several nuclei have been studied in our experimental campaign. We have measured $\beta$-delayed proton and $\gamma$ emissions and the related branching ratios. Decay schemes and absolute \textit{B}(F) and \textit{B}(GT) strengths have been determined. For some of the cases the mass excesses have also been deduced. For some of the \mbox{$T_z=-2$} nuclides under study \cite{Orrigo2014, Orrigo2016, Orrigo2018} it was possible to enrich the $\beta$-decay data by comparison with complementary ($^3$He,\textit{t}) CE reactions performed on their stable mirror target at RCNP Osaka \cite{Fujita2011, HFujita2013, Ganioglu2016}.

Some common features emerge when looking at nuclei with the same $T_z$ value. The decay of the \mbox{$T_z=-2$} nuclei proceeds by both $\beta$-delayed proton emission and $\beta$-delayed $\gamma$ de-excitation. An exotic feature that we have observed in all the \mbox{$T_z=-2$} systems studied ($^{56}$Zn, $^{48}$Fe, $^{52}$Ni and $^{60}$Ge) is the competition between the $\gamma$ de-excitation and the proton emission from the \mbox{$T$ = 2}, 0$^+$ isobaric analogue state (IAS) populated in the daughter nucleus \cite{Orrigo2016,Dossat2007}. The $\beta$-delayed proton emission from the IAS is isospin-forbidden, but it is observed and this is attributed to a \mbox{$T$ = 1} isospin impurity in the IAS wave function. In the cases of $^{48}$Fe and $^{52}$Ni the $\beta$-proton component constitutes only 14\% and 25\% of the respective IAS total decays. A reason for this is the relatively low energy of the protons emitted by the daughter nuclei, corresponding to calculated proton half-lives of the same order-of-magnitude as the $\gamma$-decay Weisskopf transition probabilities \cite{Orrigo2016}. 

\begin{figure}[ht]
\centering
\sidecaption
\includegraphics[width=1\columnwidth]{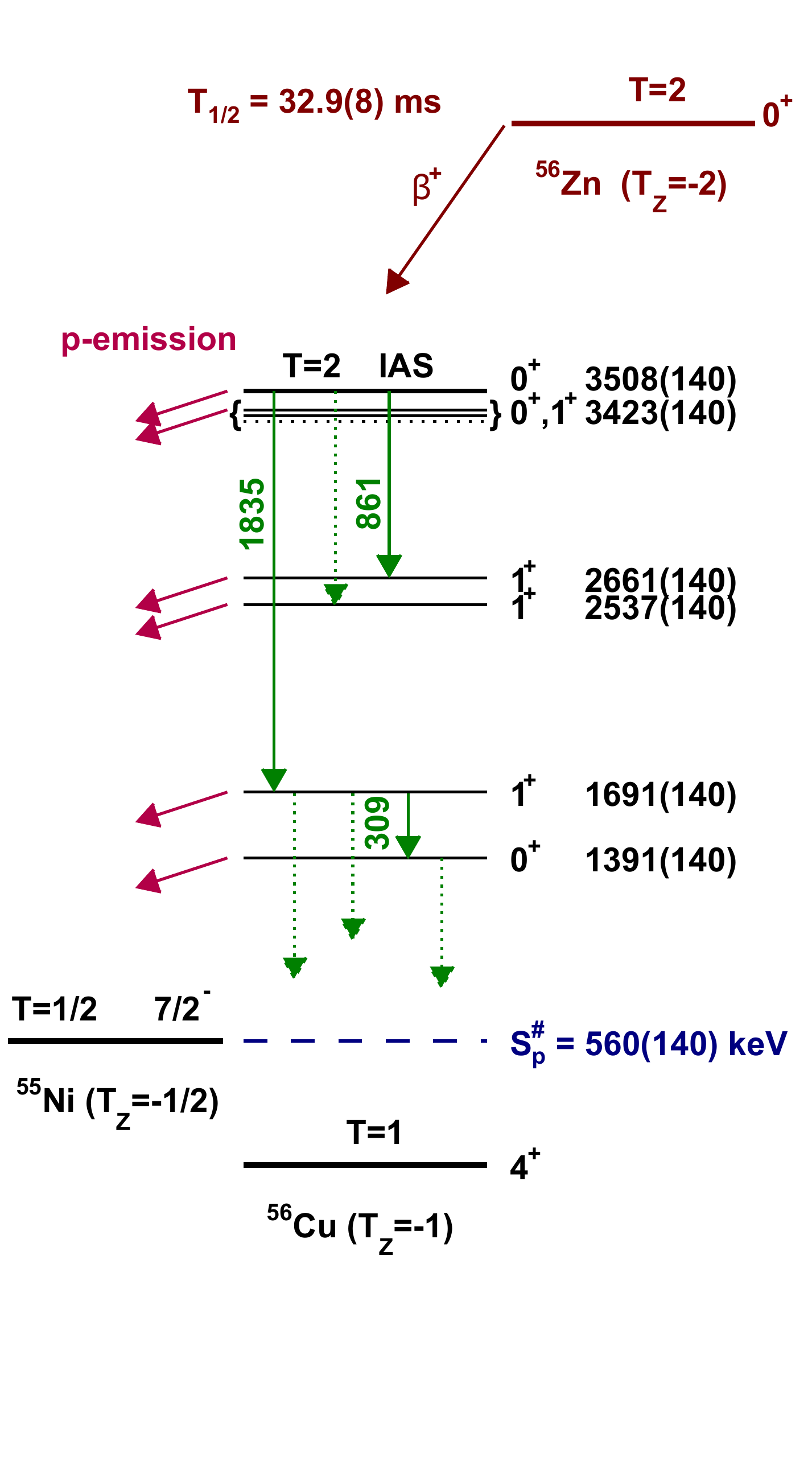}
\caption{Partial decay scheme of $^{56}$Zn (reprinted without changes from Ref. \cite{Orrigo2018} under CC BY 3.0). Transitions corresponding to those observed in the mirror $^{56}$Co nucleus are represented by dotted lines.}
\label{fig4}
\end{figure}

In the decay of $^{56}$Zn (shown in figure~\ref{fig4}), the $\beta$-proton component de-exciting the IAS is 44\% of the total, i.e., much larger. The $^{56}$Zn daughter, $^{56}$Cu, has another 0$^+$ state within 100 keV of the IAS and, since the mixing depends on how close the two states are, the strong isospin mixing of 33\% favours proton decay \cite{Orrigo2014}. At this point one would expect that the much faster proton decay \mbox{($t_{1/2}\sim~10^{-18}$ s)} from the $^{56}$Cu IAS should dominate over the $\gamma$ de-excitation ($t_{1/2}\sim~10^{-14}$ s in the mirror). However the competing $\gamma$ decay from the IAS is still observed, being 56\% of the total. The reason for such behaviour lies in the nuclear structure. Two independent shell model calculations \cite{Rubio2016, Smirnova2016} found that the proton decay of the \mbox{$T$ = 1} IAS component is hindered by a factor of 10$^{3}$. Finally, in the case of $^{60}$Ge the $\beta$-delayed proton emission from the IAS is estimated to be within 74.5\% - 95\% of the total \cite{Orrigo2021} which may again be due to nuclear structure.

\begin{figure}[ht]
\centering
\sidecaption
\includegraphics[width=1\columnwidth]{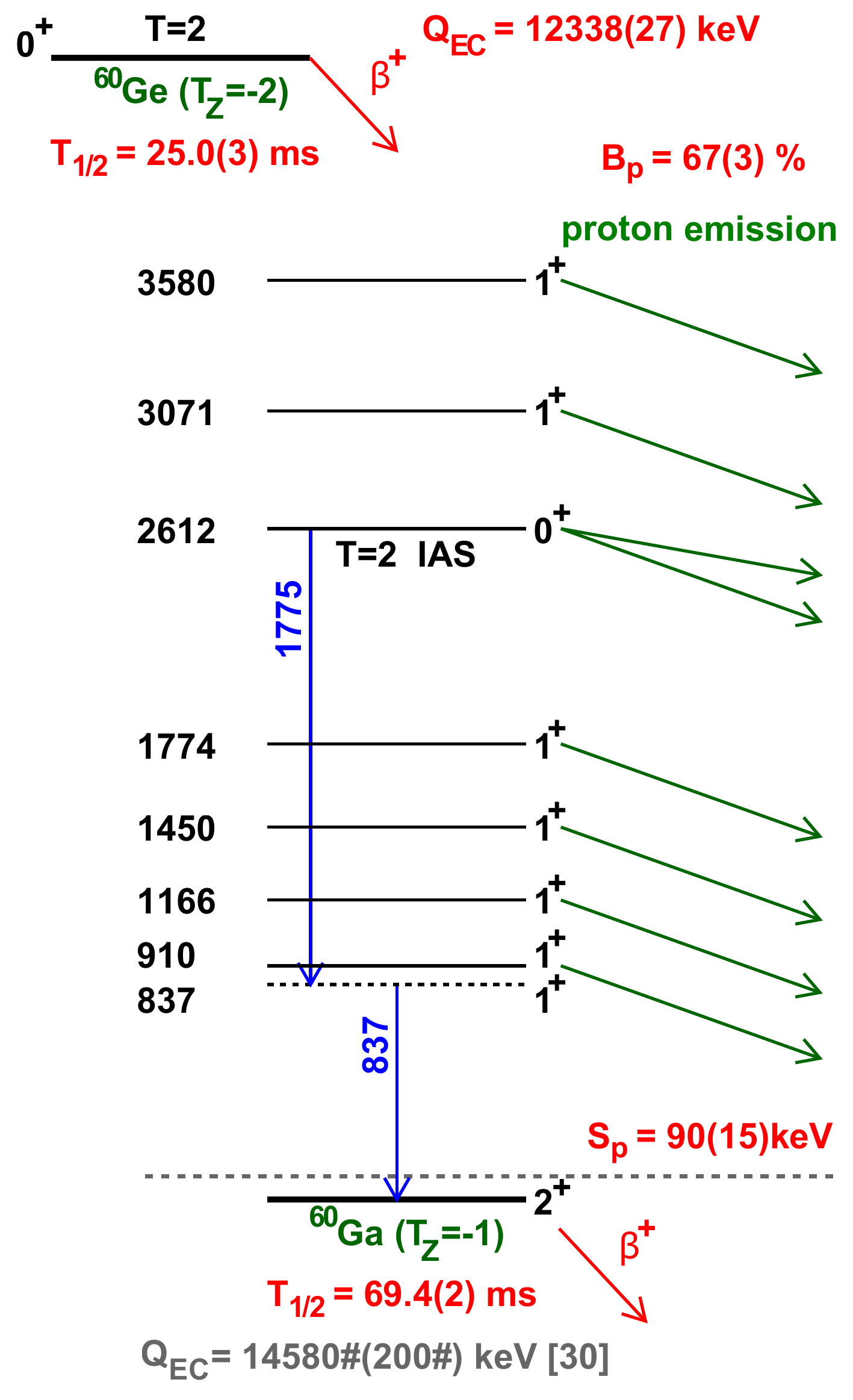}
\caption{Partial decay scheme of $^{60}$Ge. The transitions populating levels in $^{60}$Ga are shown. Observed $\beta$-delayed protons and $\beta$-delayed $\gamma$ rays are indicated \cite{Orrigo2021}.}
\label{fig5}
\end{figure}

$^{56}$Zn, lying at the proton drip-line, has been one of the most intriguing cases since it presents many interesting and unusual features \cite{Orrigo2014}. As mentioned above, the IAS is divided between two mixed levels. These levels are both fed by the Fermi transition, so that the Fermi strength is shared between them. Moreover, in $^{56}$Zn we have discovered a new exotic decay mode in the \textit{fp}-shell: the $\beta$-delayed $\gamma$-proton decay. This exotic decay pattern is possible because in $^{56}$Zn the $\beta$-delayed $\gamma$ rays populate levels in the daughter nucleus which are located above the proton separation energy and hence are unbound. The levels then decay by proton emission to the ground state (g.s.) of $^{55}$Ni. Therefore the sequence is $\beta$-delayed $\gamma$-proton decay and we have observed three such sequences. The comparison between the $\beta$ decay of $^{56}$Zn to $^{56}$Cu \cite{Orrigo2014} and the mirror CE process, the $^{56}$Fe($^3$He,$t$)$^{56}$Co reaction \cite{HFujita2013}, shows a remarkable isospin symmetry: all the dominant transitions are observed in both cases and with very similar strengths. Such a comparison allowed us to clarify some aspects of the level structure in $^{56}$Cu which would have remained unclear otherwise. 

$^{60}$Ge is another fascinating case. Its decay was almost unknown before the present experiment, where we obtained the first experimental information on both the $\beta$-delayed proton and $\gamma$ emissions, reconstructing a complex decay scheme involving five different nuclei \cite{Orrigo2021}. The partial decay scheme showing the transitions populating energy levels in the daughter nucleus $^{60}$Ga, which were unknown before, is reported in figure~\ref{fig5}. $^{60}$Ga lies right at the proton drip-line, thus its structural properties are of relevance for the rp-process \cite{Paul2021}. 

A common feature that we have found in the \mbox{$T_z$ = -1} nuclei \cite{Molina2015, Orrigo2021} is the suppression of isoscalar $\gamma$ transitions between \mbox{$J^{\pi}$ = 1$^+$}, $T$ = 0 states (Warburton and Weneser \textit{quasi-rule} \cite{Morpurgo58, Wilkinson69}). We have verified experimentally that it also holds in the heavier $^{62}$Ge nucleus. In addition, we did not find evidence of enhanced low-lying Gamow-Teller strength in $^{62}$Ga due to isoscalar proton-neutron pairing, confirming the findings of a previous measurement~\cite{Grodner2014}.

Finally, among the many results we emphasise the first observation of the 2$^+$ isomer in $^{52}$Co \cite{OrrigoPRC2}. There had been speculation that such an isomeric state exists but it was not observed earlier because, when one attempted to populate $^{52}$Co directly, both the g.s. and isomeric state of very similar half-lives are produced and implanted together. We have succeeded disentangling the two decays, measuring for the first time the isomer half-life [102(6) ms] and the g.s. half-life without contamination [112(3) ms]. The trick was to look at the implantation of $^{52}$Ni, whose decay process directly populates the 0$^+, T = 2$ IAS in $^{52}$Co. The latter then de-excites by $\gamma$-ray emission populating in a selective way the 2$^+$ isomeric state.

\begin{figure*}[th!]
\centering
\includegraphics[width=1.5\columnwidth]{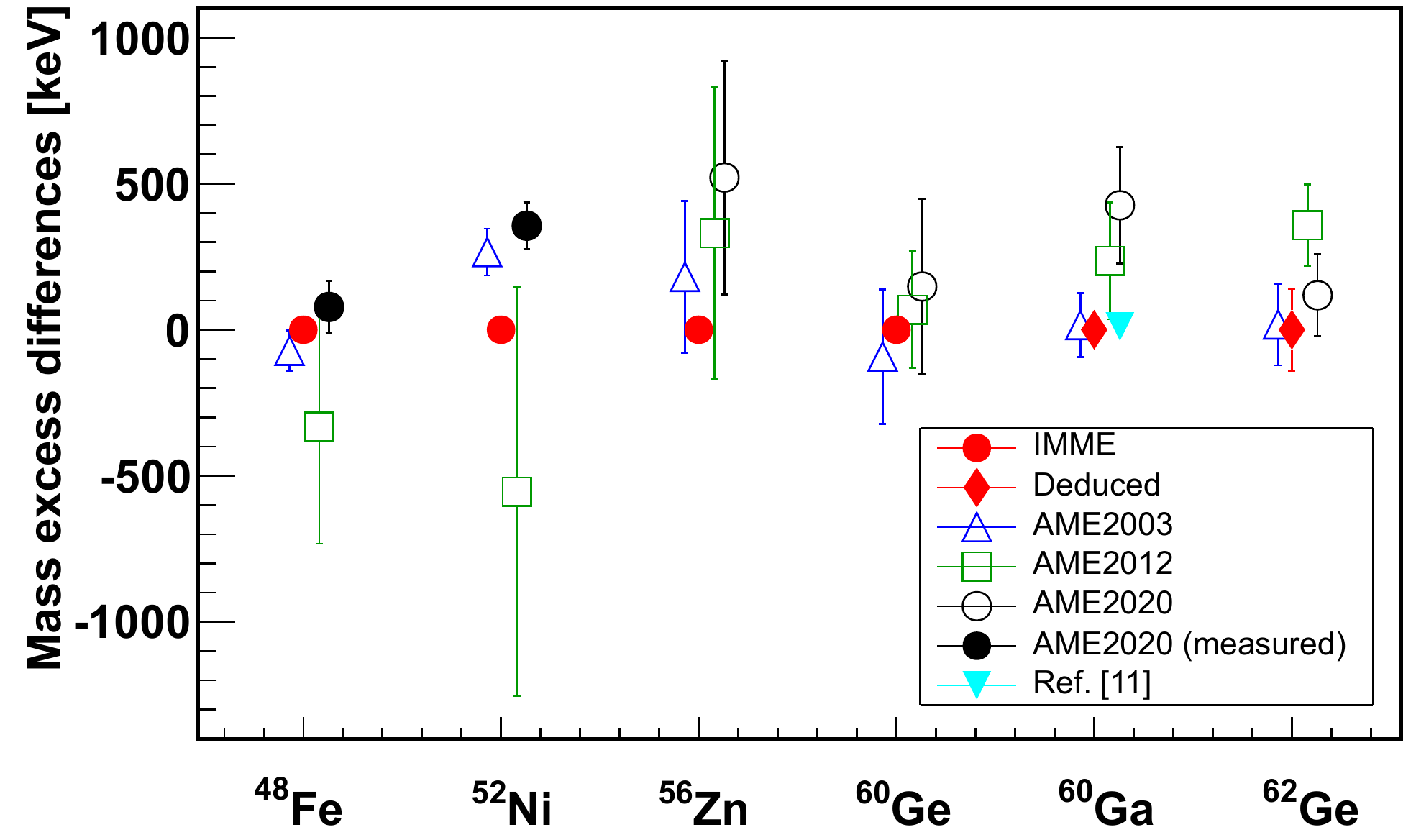}
\caption{Mass excesses of $^{48}$Fe, $^{52}$Ni, $^{56}$Zn, $^{60}$Ge, $^{60}$Ga and $^{62}$Ge. The difference is shown between the values we obtain (red filled circles and diamond) \cite{Orrigo2016, Orrigo2021} and values from the 2003, 2012 and 2020 AMEs \cite{Audi2003, Audi2012, Wang2021}. Open symbols represent values from systematics, while filled symbols are experimental or deduced values. The cyan filled triangle is the measurement from Ref. \cite{Paul2021}. The data points belonging to each nucleus are slightly displaced to show the error bars better.}
\vspace{-5 mm}
\label{fig2}
\end{figure*}

\section{Mass excesses}
\label{sec2}
The mass excesses of the $\beta$-decaying neutron-deficient nucleus and its daughter nucleus are important to determine key quantities such as the $Q_{\beta}$ value of the decay, needed to deduce the \textit{B}(F) and \textit{B}(GT) strengths. The proton separation energy \textit{S$_p$} in the $\beta$-daughter nucleus can be determined by knowing its mass excess and that of the $\beta$-proton daughter nucleus. 

There is a lack of mass measurements in this mass region. When the mass excesses are not known, the atomic mass evaluation (AME) systematics can be used. Another option is to determine the mass excess of the nucleus of interest from the Isobaric Multiplet Mass Equation (IMME) \cite{IMMEpaper, IMME, MacCormick2014}:

\begin{equation}
  \Delta m(\alpha,T,T_z) = a + b~T_z + c~T_z^2\,.
  \label{Eq1}
\end{equation}
\\
This can be done when at least three other members of the isospin multiplet are known. In equation~\ref{Eq1}, $T_z$ is the third component of the isospin $T$ and $\alpha$ stands for all the other quantum numbers. From our $\beta$-decay data we have determined the g.s. mass excesses of the \mbox{$T_z$ = -2} nuclei $^{48}$Fe, $^{52}$Ni and $^{56}$Zn in Ref. \cite{Orrigo2016} and $^{60}$Ge in Ref. \cite{Orrigo2021} from the IMME, knowing four members of each quintuplet. We have also deduced the g.s. mass excesses of $^{62}$Ge and $^{60}$Ga, and \textit{S$_p$} in $^{60}$Ga \cite{Orrigo2021}. Our deduced mass excesses (red filled circles and diamonds) are shown in figure~\ref{fig2} and compared with the values obtained from the 2003 \cite{Audi2003}, 2012 \cite{Audi2012} and 2020 \cite{Wang2021} AME systematics. In the figure open and filled symbols represent the measured and deduced values in the AME systematics. 

We have observed since our first study of $^{56}$Zn \cite{Orrigo2014} that, for proton-rich nuclei in this region of the mass chart, the AMEs published subsequent to that in 2003 are in poorer agreement for the mass excess in comparison with our IMME or deduced values. After us, other authors reported similar issues \cite{DelSanto2014, Paul2021}. As shown in figure~\ref{fig2}, the values from the 2003 AME lie closer to our estimates than the values from the 2012 AME. The values from the 2016 AME \cite{Wang2017} (not shown in the figure) have a very similar behaviour. The 2020 AME also behaves similarly for the unmeasured nuclei, but includes new measured values for $^{48}$Fe and $^{52}$Ni (black filled circles). The AME systematic evaluation does not include isobaric multiplets because the IAS might be mixed and thus its energy might deviate from the IMME formula. An IAS-mixed case is $^{56}$Zn \cite{Orrigo2014}. We think that, with caution, this knowledge can help to calculate extrapolated values. New mass measurements in this region are important to better constrain the future AME. 

A recent measurement of the mass excess in $^{60}$Ga is also shown in figure~\ref{fig2} (cyan filled triangle) \cite{Paul2021}, which is in excellent agreement with our indirect determination from the $\beta$-decay data \cite{Orrigo2021}. Paul \textit{et al.} have also determined \textit{S$_p$}= 78(30) keV in $^{60}$Ga, in agreement with our value of 90(15) keV \cite{Orrigo2021}. The 2020 AME systematic value is -340(200) keV \cite{Wang2021}. By combining the two experimental values, \textit{S$_p$}($^{60}$Ga) = 88(18) keV is obtained, establishing the proton-bound nature of $^{60}$Ga. This value, together with the fact that $^{59}$Ga was not observed in fragmentation reactions at NSCL, provides strong evidence that $^{60}$Ga is the last proton-bound gallium isotope \cite{Paul2021}. $^{59}$Ga was also not observed in our recent experiment at RIKEN, strengthening the conclusion that $^{60}$Ga marks the location of the proton drip-line for $Z$ = 31.

\section{Half-life trends}
\label{sec3}
\begin{figure}[ht]
\centering
\sidecaption
\includegraphics[width=1\columnwidth]{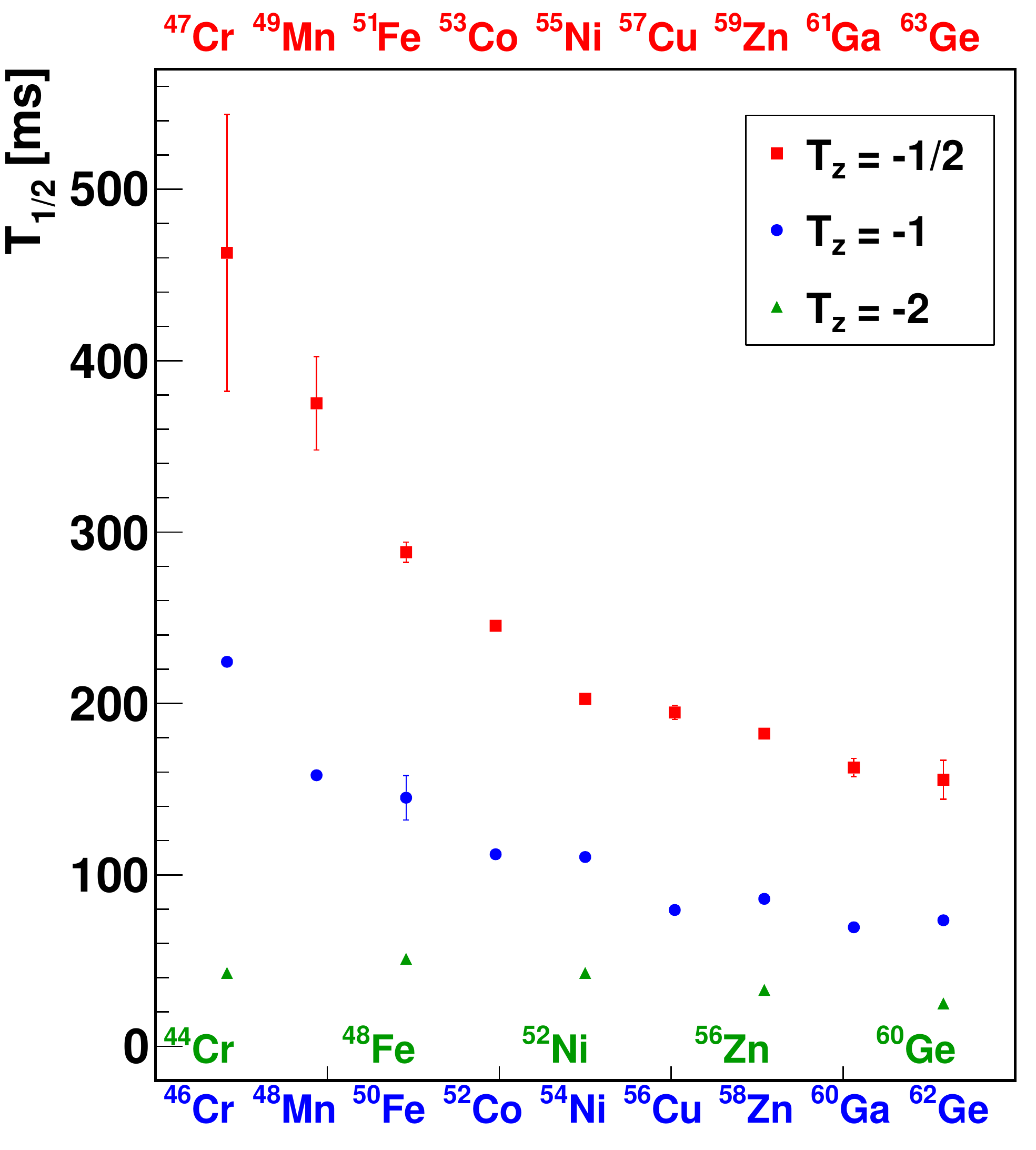}
\caption{$T_{1/2}$ values measured for the $T_{z}$ = -1/2 (red squares), -1 (blue dots) and -2 (green triangles) nuclei as a function of atomic number.}
\label{fig3}
\end{figure}

In the present Section we summarise the results concerning the half-lives of the nuclei studied and analyse their trends. Figure~\ref{fig3} represents all the measured half-lives as a function of the atomic number $Z$ and for the $T_{z}$ = -1/2, -1 and -2 nuclei. It is an updated version of figure 7 from Ref. \cite{Kucuk2017}, where we have included/updated the half-lives of the following nuclides: $^{56}$Zn \cite{Orrigo2014}; $^{48}$Fe and $^{52}$Ni \cite{Orrigo2016}; $^{52}$Co \cite{OrrigoPRC2}; $^{60}$Ge, $^{60}$Ga, $^{62}$Ge and $^{59}$Zn \cite{Orrigo2021} and $^{44}$Cr \cite{Dossat2007}. 

Three curves are obtained, corresponding to the different $T_{z}$ values, because the $B$(F) value driving the decay is the same for all the nuclei with the same $T_{z}$. As discussed in Ref. \cite{Kucuk2017}, the systematic decrease of the $T_{1/2}$ values with the mass, seen in the $T_{z}$ = -1/2 nuclei, reflects the increase in the $Q_\beta$ value. A similar decreasing pattern is observed in the half-lives of the $T_{z}$ = -1 nuclei and, on top of this behaviour, a typical odd-odd and even-even effect is found. The latter is due to the fact that in the odd-odd nuclei there exist other excited states below the IAS that receive a significant amount of $\beta$ feeding, which makes their half-lives slightly shorter in comparison with their even-even neighbours \cite{Kucuk2017}. In the $T_{z}$ = -2 nuclei, which are all even-even, a smoother decreasing trend is observed. The half-life of the most exotic $T_{z}$ = -2 nucleus, $^{60}$Ge, is only 25.0(3) ms. 

\section{Conclusions}
\label{conc}
We have given an overview of the most relevant achievements from our $\beta$-decay spectroscopy experiments at GANIL and RIKEN. Detailed spectroscopic information has been obtained for several neutron-deficient nuclei, starting from lighter to heavier systems along the proton drip-line. Half-lives, decay schemes, $\beta$-decay transition strengths, mass excesses have been determined, many of them for the first time. These results are relevant for both nuclear structure and nuclear astrophysics.

Our deduced mass excesses have been compared with different AME systematics, indicating the need for more mass measurement in this region of the nuclear chart. The half-life values as a function of the mass number have been analysed, providing a comprehensive understanding of the half-life trends in terms of the Fermi strength and the $Q_\beta$ value.

We have shown that valuable spectroscopic information can be obtained from this kind of experiment, thus improving our knowledge of the properties of neutron-deficient nuclei.
\\

\begin{acknowledgement}
\textbf{Acknowledgments}. This work was supported by the Generalitat Valenciana Grant No. PROMETEO/2019/007, Spanish Grants No.~PID2019-104714GB-C21, FPA2017-83946-C2-1-P and FPA2014-52823-C2-1-P (MCIN,MINECO/AEI/FEDER), Centro de Excelencia Severo Ochoa del IFIC SEV-2014-0398; \textit{Junta para la Ampliaci{\'o}n de Estudios} Programme (CSIC JAE-Doc) co-financed by FSE.
\end{acknowledgement}
%
%
%

\end{document}